# Process Patterns for Service-Oriented Development

Mahdi Fahmideh[1], Mohsen Sharifi[2], Fereidoon Shams[1], Hasan Haghighi [1]

[1] Automated Software Engineering Research Group, ECE Faculty, SB University GC, Tehran, Iran
{m.fahmideh, f_shams, h_haghighi}@sbu.ac.ir

[2]School of Computer Engineering, IUST, Tehran, Iran
msharifi@iust.ac.ir

*Abstract*—Software systems development nowadays has moved towards dynamic composition of services that run on distributed infrastructures aligned with continuous changes in the system's requirements. Consequently, software developers need to tailor project specific methodologies to fit their methodology requirements. Process patterns present a suitable solution by providing reusable method chunks of software development methodologies for constructing methodologies to fit specific requirements. In this paper, we propose a set of high-level service-oriented process patterns that can be used for constructing and enhancing situational service-oriented methodologies. We show how these patterns are used to construct a specific service-oriented methodology for the development of a sample system.

I. INTRODUCTION

The subject of Service-Oriented Computing (SOC) spans many concepts and technologies that find their origins in diverse disciplines and are interwoven in a complicated manner [1]. SOC paradigm has been inspired when developing such software applications at organization level and ultra large-scale levels namely Systems of Systems (SoS). Therefore, software engineering practitioners and researchers continue to face huge challenges in the development and maintenance of service-oriented software systems. This has been even more prominent when development teams need to create methods, tools, and techniques to support cost-effective development and use of diverse services to construct a service-oriented system.

SOC research studies have stated crucial challenges and concerns in the development of service-oriented systems [1,2]. For instance, what activities should be conducted for the development of service-oriented system? How to modernize legacy systems towards service-oriented system? What organizational/SoS issues are incorporated in the development of these type of systems? What activities should be conducted on business processes to attain suitable services? What is required is a service-oriented software development methodology that accommodates these challenges. In this regard, various methodologies have emerged to support the lifecycle of service-oriented development. Based on the unified definition of methodologies [3], and the various reports on analyzing existing service-oriented methodologies [4,5], the most prominent and popular ones are as follows: IBM SOAD [6], IBM SOMA 2008 [7], CBDI-SAE Process [8], SOUP (Service-Oriented Unified Process)[9], MASOM (Mainstream Service-Oriented Architecture Methodology) [10], SOA RQ [11], Papazoglou [12], RUP for SOA [13], Service-Oriented Architecture Framework (SOAF) [14], Steve Jones' Service Architectures [15]. The main reasons for selecting these methodologies were maturity level, number of citation, adequate resources and proper documentation. However, authors recognize that there are two key challenges in them that should be addressed:

- **Weaknesses of the acclaimed service-oriented methodologies**: None of the existing service-oriented methodologies covers all issues of service-oriented development; they are only pertinent to specific aspects of service-oriented development [4,5]. Three weaknesses have been identified: 1) lack of full coverage of service-oriented development life cycle (SOAD for instance), 2) lack of supportive documents on their practical use (SOMA 2008, CBDI-SAE, RQ and SOUP for instance), and 3) cursory development process models (SOUP for instance).
- **Multiplicity of notions**: Each service-oriented methodology supports different activities. Interestingly, most of the methodologies prescribe different activities with different names that are in fact similar. They have the same activities but from a different viewpoints. In an abstract view, we can find out recurring activities in their development processes. Multiplicity and similarity of the service-oriented methodologies confound the users to select an appropriate one.

In this regard, general service-oriented methodologies can resolves these challenges through addressing the shortcomings while being adjustable according to the details of the project situation at hand. The need for developing situation-specific methodology has led to *Situational Method Engineering* (SME) [16] wherein a project-specific methodology is constructed from *reusable method chunks*. Specifically, *assembly-based* approach of SME uses reusable method chunks of existing methodologies to construct project specific methodologies by selecting and assembling method chunks obtained from different methodologies that are stored in a library [17]. For constructing service-oriented specific methodology, a number of comprehensive sources of method chunks that inspired by service-oriented context are needed. To obtain this source, one suitable candidate is *Process Patterns* [18]. Process Patterns are classes of common successful practices and recurring activities in methodologies is represented [19]. They are result of applying abstraction to successful software development methodologies that form a process meta-model of software development. In this regards, process patterns can thus useful to provide a library as method chunks so that method engineer can select most appropriate patterns that satisfy the context requirements and construct a new methodology through assembling them.

Recently researchers have proposed process patterns in a different context of software development. The OPEN (Object-oriented Process, Environment, and Notation) Process Framework (OPF) is a set of process patterns used for constructing project-specific object-oriented methodologies [20]. Other researches have been conducted for defining Agile development process patterns (Agile Software Process) [21] and decision support software development process patterns [22]. Since no contribution in the service-oriented development had been found, we therefore propose a comprehensive set of *Service-Oriented Process Patterns*, called SOPP, commonly encountered in prominent service-oriented methodologies as the source so that they can be used in constructing project-specific service-oriented methodologies. Process patterns can be imported as plug-ins into software process management environments such as Eclipse Process Framework (EPFC) [23] or Rational Method Composer (RMC) [24].

The rest of this paper is structured as follows. Section II presents SOPP in detail. Section III provides an illustrative example of applying SOPP to a real-world service-oriented project. In section IV discusses how the proposed process patterns can be evaluated. Finally, Section V concludes the paper.

## II. PROPOSED SERVICE-ORIENTED PROCESS PATTERNS

Although process patterns are commonly recurring activities in the software development methodologies, we need a technique to allow method engineers to analyze activities of different methodologies in order to extract meaningful process patterns. To do this, we have adopted the technique previously introduced in [25]. We represent our extracted process patterns in terms of *Problem, Context, Solution, Typical Roles* and *Artifacts* [19], and organize patterns in a cohesive generic process meta-model in three levels of abstraction based on granularity of patterns. The resulted repository contains 4 *Phases*, 9 *Stages* and 49 *Tasks*. A *task process pattern* defines the required steps to execute as task (e.g. technical code review). *Stage process pattern* are contained several tasks process patterns that need to be done to pass from a stage of development. Typically, they perform in iterative-incremental manner (e.g. Design Architecture). Two or more stage patterns form a *Phase* patterns as the typical phase of software development life cycle. Moreover, for each pattern, input/output artifacts and typical general roles are assigned. There is no predefined constraint to run stage patterns in successive order, unless method engineer in a specific methodology concretes them. SOPP focuses on specific patterns dealing with specific concerns of the service-oriented development methodologies. Due to space limitation, we ignore to re-explain general repetitive activities as patterns that are mandatory in any methodology such as risk management, requirements management, service change management and versioning, project management, distributed team management and alike that exist in any typical software process.

### A. Phase Process Patterns

By applying the proposed technique for extracting process patterns from ten prominent service-oriented methodologies, we identified four main phases as follows: 1) *Initiate*, 2) *Design Service-Oriented Solution*, 3) *Assemble*, and 4) *Maintain* (Fig 1). The *Initiate* phase pattern contains activities for elicitation of high-level requirements and analysis of the state of existing organizational/SoS situations. The objective of *Design Service-Oriented Solution* phase is to justify of candidate valuable steps of business processes and expose them as services to external consumers. A set of primary high-level services is then defined based on the business processes, designed services fall into instantiated SOA stack. Overall, the phase maps business processes to a set of services comprising a service-oriented architecture.

The *Assemble* phase develops required services and integrates them to form a service-oriented software system. There are alternatives to service development: some of the services are provided by wrapping of existing legacy systems' interfaces to provide a coarse-grained service, or may be purchased from external service providers. Otherwise, if a group of required services that cannot be provided in none of the above alternatives, should be constructed from scratch.

Due to the nature of service-oriented context, a new service-oriented software system may be fully constructed through assembling a number of existing independent services. Therefore, it is reasonable that *Construct Services* and *Test Services* stages to be omitted in some situations. In the *Maintain* phase, the services' interfaces are published, added to a service repository and made operational. The phase is also concerned with preserving high quality of services (QoS) in the operational environment. All patterns are not mandatory in the development a service-oriented project, so a method engineer can select those that are appropriate for the project at hand.

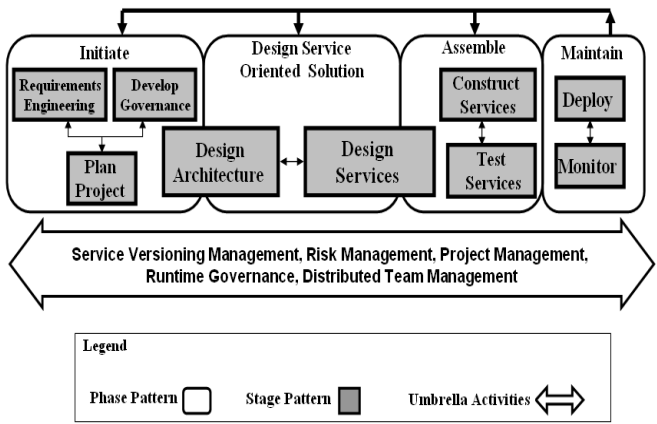

Fig. 1. SOPP as a generic service-oriented process meta-model

### B. Stages and Task Process Patterns

In this section, we describe the details of stage and task process patterns. Some specific details of how to perform a task pattern are specified since they are inspired by various techniques have already been proposed in the service-oriented methodologies or other context. For instance, there are three supportive techniques for *Service Definition* task pattern namely *Top-Down*, *Bottom-Up* and *Meet-In-the Middle* [7,10]. In this regards, method engineer can assigns these techniques to task pattern. Therefore, we will not discuss task process patterns in detail. In further in section 1 to 6 we present stage process patterns in detail.

*1) Requirements Engineering*
**Problem**: What are the requirements of a service-oriented system?
**Context**: Organizations either independently or collaboratively want to expose their business processes as services to external consumers.
**Solution**: As shown in Fig.2, the requirement engineering stage pattern for service- oriented software development is very similar to traditional requirements engineering. Differences are in *Specify SLA* task pattern, *Analyze Environment* and *Analyze Business Processes* stage patterns that are common in service-oriented methodologies. In *Specify SLA* relevance qualities relating to services (business services or application services) such as security, performance and so on are specified. In the *Analyze Environment* the status of existing infrastructure of environment is analyzed to figure out the amount of efforts needed for migration and also to build proper services from the existing assets and evaluates the readiness of environment to migrate to service-oriented solution. Moreover, the reasons of migration to service-oriented are justified. The *Analyze Business Processes* is performed when optimization to business processes is required. Also, it identifies related supplemental business rules and constraints to business processes.

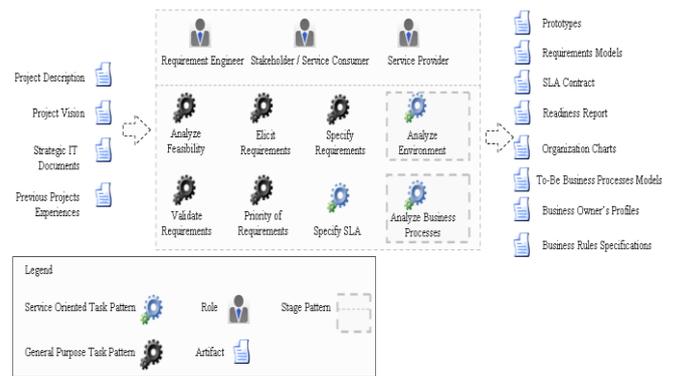

Fig.2. Requirements Engineering stage process pattern

*2) Develop Governance*
**Problem**: How to ensure that the right services are developed and are aligned with the environment strategies and business process goals? How to ensure that a stable and suitable collaboration between business stakeholders and development team are maintained during software development?

**Context**: The project has started, however, imperative environment policies and mechanisms to mitigate service-oriented pitfalls and prevent falling into wrong direction should be defined.
**Solution**: In this stage a governance model is established and applied to the whole development lifecycle (Fig. 3). Governance model specifies the policies, rules, procedures and measurement metrics to ensure that software development, as a set of services, are constantly aligned with IT initiatives. Services should be traceable back to business objectives. In the *Plan Governance for Project Iterations*, stakeholders collaborate to establish a scope and fund for performing the governance model in current iteration. *Define Policies and Procedures for Criterion* defines a set of supportive policies and rules to achieve right services that essentially relate to quality attributes. For this purpose, metrics and indicators are defined to measure and monitor quality of services by *Define Indicators and Metric for Measurements* task pattern. In the *Enable* task pattern, the governance model is rolled out, published, and monitored to various stakeholders across environment.

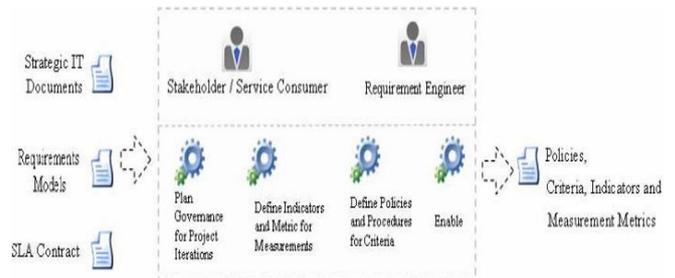

*3) Design Services*
**Problem**: How business processes are exposed through business and application services to available external consumers?

*Context*: A set of selected environment business processes that have been represented and mostly improved. They are prepared for transformation to a set of services.

*Solution*: The *Design Services* stage is the core of the SOPP meta-model (Fig. 5). When business processes in the focal area of environment are identified and re-engineered, useful services that encapsulate business capabilities should be defined. This stage takes a set of refined business process models as input and yields a set of candidate services. Firstly, business processes are translated into one or more services by performing *Define Service* task pattern. Having defined services, initial interfaces are created. These interfaces are refined by modeling and analysis of their collaborations (*Analyze Service Collaboration*). The *Specify Service Interface* task pattern is responsible for consolidating the interface with more specific details such as interface signatures, operation parameters, protocol information and input/output messages. The aim of the *Evaluate Services* is to simplify the maintenance and future enhancements of services. In this task, quality of designed services such as Determining Right Level of Granularity, Degree of Cohesion and Coupling and Reusability are checked. In the *Classify Services*, services are placed in service groups based on their usage context such as mission-aligned business services, application services or common enabling IT services such as authentication, authorization, logging, notification and utility services. Finally, the *Design Database* designs a required repository for persistence data storage. For instance, utility services are required to record message details in a notification log database. Furthermore, dependencies of services on the current version of the real databases must be carefully sorted out. As it was shown in Fig. 1, this stage lies in the *Develop* phase, because the services prepared in the stage may be refined again when services are elaborated during development.

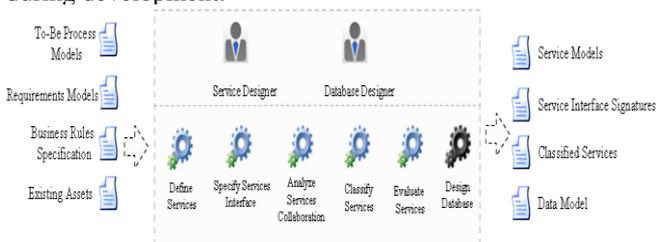

Fig.3. Design Services stage process pattern

### 4) Construct Services

*Problem*: How to construct the designed services?

*Context*: Detailed design of services has been produced.

*Solution*: In the *Construct Services* stage (Fig. 7), the required services are developed. The *Choice Implementation Alternatives* evaluates existing alternatives to obtain services. Based on the service models and existing software assets, some of the services are provided by wrapping important functionality via universal standard Web Service technologies (*Implement Wrappers*).

A group of services are provided by discovering published services in the Internet or commercial companies (*Discover Services*). If services are acquired in this way, *Certificate Service* is performed to ensure they satisfy the required quality of concerns (SLA) according to certification before using them in the system. In the case of no exact match with the requirement(s), appropriate service(s) are developed by the development team (*Implement Services*). Indeed, in case of implementation of services, a well-known object-oriented analysis and design techniques such as use-case based analysis, grammatical parsing and CRC card modeling can be accommodated for identifying, analyzing and designing software classes. The relevant classes are then classified into a number of cohesive software components and consequently cohesive software components are classified to form Services.

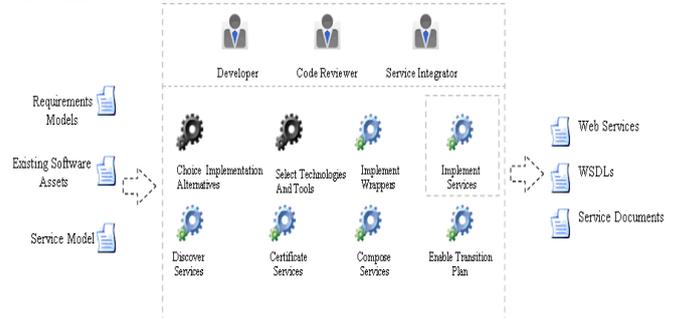

Fig. 4. Construct Services stage process pattern

In the *Choice Technologies and Tools*, the development team selects appropriate technologies and tools (such as Microsoft .Net, J2EE or BPEL) for developing the system. Moreover, provided services can be combined to realize expected composite business services to provide complex business processes (*Compose Services*). A complex business process can be built based on existing independent business services. At runtime, each service can be replaced with another one with respect to high-level policies, performance issues, SLA stipulation, and so on (Section 2.2.9: *Compose Service Dynamically* task pattern). Alongside constructing services in different manners, the *Enable Transition Plan* is performed for the purpose of modernizing legacy systems with service-orientation as well as previously defined strategic plans for moving towards a service-oriented solution.

### 5) Deploy Service

*Problem*: How are the provided services deployed on computing platforms?

*Context*: System, as a set of composed independent services, is ready to be deployed on computing platforms and the whole system is validated.

*Solution*: In this stage, services as building blocks of the system are ready to be deployed in an operational environment in which the service-based system should become available to service consumers (Fig. 9). Moreover, defects and missing requirements are discovered in this stage. The *Prepare Infrastructure* prepares the necessary network, software and hardware platforms of the architecture as it is defined in the *Design Service-Oriented Architecture*, before the deployment of services. In the *Publish Services*, final services are deployed by service providers. Furthermore, they are added to common services directory (UDDI protocol) to allow service consumers to discover the existence and location of services (*Add to Services Repositories*). In the "*Test Orchestrations/Choreographies Of services*", the system is tested to see if the composition of services that build the system actually meet the business acceptance criteria for functional requirements and SLA for nonfunctional concerns on a distributed network. A service-oriented system may be developed simply by composing or orchestrating a set of existing services that have already been available on an accessible distributed network, without requiring much effort in their deployment.

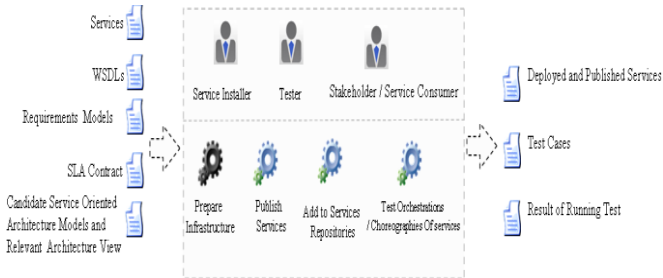

Fig.5. Deploy stage process pattern

*6) Monitor Services*

**Problem**: How to constantly ensure the health and quality of services during system execution?

**Context**: All services have been fully deployed and become operational by the development team (as service providers) and service consumers invoke service operations.

**Solution**: After system is fully deployed in an operational environment, the *Monitor Services* stage evaluates functionalities and QoS of the operational system continuously (Fig. 10). In the *Monitor SLA* task pattern, the quality of services is measured and analyzed by gathering and logging data during service usage by service consumers. The development team performs this task to address issues raised by rectifying noncompliance with functional requirements and service qualities, before failures actually occur. Even in cases where a service-based solution is constructed by the composition of existing independent services, malfunctioning of services at runtime or any violations from the SLA agreement in the system must be sensed and rectified dynamically by replacing some services with other services (i.e. reconfiguration by the *Compose Services Dynamically* task pattern).

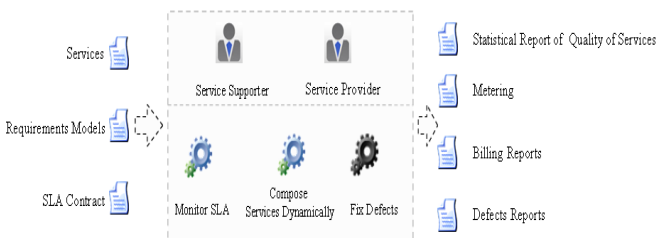

Fig. 6. Monitor stage process pattern

## III. ILLUSTRATIVE EXAMPLE: APPLICATION OF SOPP IN DESIGNING A PROJECT SPECIFIC SERVICE-ORIENTED METHODOLOGY

In this section, we illustrate the application of the proposed process patterns; how it can be used to construct a specific service-oriented methodology for the development of a sample system. We first elicit the methodology requirements, and then design it by instantiation and adaptation of the SOPP. The case study is about developing a service-oriented system for providing some residential services to employee of an NGO [27]. The NGO has offices in 30 provinces with a total number of 14000 employees. Based on the business process viewpoint, the system should provide online services for booking a room and accepting the payments for the expenses. After deploying the system, any employee can send his request to book a room in one of the hotels located in a specific province and track his/her request and pay the expenses by online services provided by third party payment services. Having received the requests, the priorities are automatically determined by the system and a room is assigned to the employee. The system notifies the employee by email and SMS services and the employee confirms the reservation process. The aim is to satisfy the methodology requirements via appropriate process patterns leads to design the required methodology.

Efforts aiming at developing methodologies should begin by clearly defining what the situational requirements of such a methodology are. Method engineer is responsible to map the elicited high-level requirements of the project to process patterns that have formed a method library. For simplicity, we accommodated direct map between methodology requirements and process patterns as recommended in OPEN/OPF [20] instead of approach that has been offered by Ralyte as *Requirements Map* [28]. When the methodology requirements were fixed, method engineers clarified methodology requirements as shown in Table I. Some of the requirements were imposed by stakeholders. For instance, business processes modeling and improvement was forced due to the explicit request of stakeholders to receive a detailed documentation of their as-is and to-be business process. Other requirements were relevant to methodology quality such as agility of development process, fast responsiveness to business volatility, flexibility, time and cost of system development.

To realize methodology requirements through SOPP, method engineer started by setting the overall life cycle at the highest level of abstraction by using the *Initiate*, *Define Service Oriented Solution*, *Assemble* and *Maintain* phase patterns. After that, method engineer elaborated the methodology using stage patterns and subsequently their task patterns. To do this, method engineer took a set of consecutive decisions based on the requirements and their relations with patterns. By considering the list of *Problem/Context/Solution* section of each pattern, method engineer figured out which pattern(s) match a requirement (denoted by M).

For instance, method engineer selects the *Specify SLA*, *Discover Services*, *Certificate Services*, and *Monitor SLA*

task patterns to satisfy #R1. For improving existing business processes, method engineer selects *Analyze Business Process* stage pattern to explore organization business processes and re-engineered them base on need (refer to #R2 and #R3). While a number of residency systems had been developed independently in the organization and now they became obsolete the *Analyze Environment* task pattern is selected to assess the documents of the legacy systems whether has any underlying asset to be reused (refer to #R4). The pattern had significant effect on reducing cost and time of development.

Moreover, the old residency system's databases have large amount of history records about employees. They should be available in the new system without losing the integrity. In this regard, the *Plan Transition* and *Enable Transition Plan* are selected (refer to #R5). As the last functional requirement that the custom methodology should be supported, the *Monitor SLA* and *Compose Services Dynamically* are selected to satisfy #R6. For instance, another one will replace e-bank services while the availability of current service provider is reduced. Furthermore, while the organization have plan for migration to Service-Oriented Architecture method engineer selected the *Design Architecture* (refer #R8). Selection of some patterns is unavoidable due to special situation of the project (denoted by D). For instance, the selection of the *Define Service* task pattern is obviously due to defining and exposing residency business processes as services. Another instance is requirements engineering stage patterns and its task patterns.

Now the overall development process has been instantiated from SOPP via selection of appropriate phases, stages and task patterns. But yet it shows "*What*" activities as task patterns that should be performed rather than "*How*" to be performed. More details of how task patterns instructed should be defined explicitly. Task patterns can be concretized through various supportive techniques. As shown in table II for each selected task pattern method engineer has associated a specific technique.

Some of the techniques have taken from exiting service-oriented methodologies. For instance to define what right business services should be defined method engineer associates *Top-down* and *Bottom–up* approaches as well as suggested in MOSAM to Define Services task pattern [10]. Furthermore, in according to the substances of #R7, #R9, #R10 and #R11, method engineer should learn how to utilize *Agile Methodologies* [29] in order to provide agility of development process. In this regard, method engineer utilized "Define User Story", "Evolutionary Prototyping" and "MoSCoW Rules" techniques from XP and DSDM methodologies to instantiate and concretize the task patterns in support of agility. "User Histories" captures essential functional requirements of the system and has little stress on documentation (#refer to #R9). "Evolutionary Prototyping" demonstrate expected functionalities that are iteratively refined during system development. Finally, for some of the task patterns existing general techniques have been adopted (denoted by EGT) which are most commonly used in any situation and so incorporated in constructed methodology.

Table II shows how each requirement of the designed methodology is traced back to the process patterns and to the incorporated techniques. There is no imperative one-to-one relationship between the requirements and the task patterns. Responsible roles and related artifacts are not shown.

The important point to note is that the resulting methodology must be further refined and adapted iteratively by the method engineers during maintenance of the system in accordance with project situation through iterative process reviews of the development process.

Table I. Methodology Requirements

| Key | Name | Explanation |
|---|---|---|
| #R1 | Utilizing external services | Organization decided to use third party e-bank services to supply chain of business processes. |
| #R2 | Managing frequently changes in business processes | Business rules for accepting or rejecting a request by employee will be changed frequently. Flexible adjusting of business rules and parameters should be addressed. |
| #R3 | Improving business process | The improvement of residency business processes was imperative. |
| #R4 | Using legacy systems services | In order to reduce cost and effort of system development, potential legacy functionalities should be reused. In this regard, an old Fox Pro resident program existed irrespective of being out of date. |
| #R5 | Modernizing legacy systems | Existing NGO legacy system and related operational databases should be modernized without stopping the current business processes. Traditional databases should be replaced by novel technologies. |
| #R6 | Conforming to stated quality of services | Quality of external services, specifically full availability and rate of discount per transaction are essential requirements. |
| #R7 | Agility | Faster execution of development process via the production of essential documents. |
| #R8 | Stack-based SOA (Architecture-Based) | In order to have successful IT transition plan, different system should be developed to expose valuable services to external consumers. Need of a supportive infrastructure for exposing services to fulfill authorized consumers is felt. |
| #R9 | Requirements-based | Elicited requirements should be considered in development of services and consequently target system. A past unsuccessful experience in NGO domain has shown a miss-understanding of requirement is made to develop a useless system. |
| #R10 | Project size | Expected system functionalities are limited estimably to maximum of 15 use cases. |
| #R11 | Team size | Development team is limited to ten members. They have experience with XP programming and Agile principles such as pair-programming, test-base development and evolutionary prototyping. |

## IV. DISCUSSION

To achieve a "True Assessment" of this research we aim to open two discussions in this regard. Firstly, we have proclaimed in the research that we have provided a set of process patterns for service-oriented development principally; each successful and mature service-oriented SDM prescribes best solutions in term of required activities, guidelines and supportive techniques, roles, and required artifacts. A method engineer can consider them collectively and extract a set of similarities and recurrent successful activities. We believe that original service-oriented SDMs have already attested the suitability and applicability of recurrent activities, or better say method fragments. We have extracted proposed process patterns from a number of existing service-oriented SDMs. These patterns only capture recurrent pre-examined best practices. As consequent, applicability of the proposed method fragments has been verified.

Secondly, a true empirical assessment is vital to demonstrate how the new process patterns can be utilized in more real software projects and how they increase quality of development process in term of speed and cost of software development. It should be noted "Software Process Assessment" is remained as a notoriously challenge in the SME literature and few real case studies can be found in order to industrial usages [30]. The proposed process patterns fragments give first cut of service-orientation concerns should be incorporated during service-oriented development and situational methodology construction. The software development organization, especially method engineers, should maintain, improve as time progresses and while getting feedbacks from software development teams continuously. As a concrete solution to achieve true assessment of the process patterns adopting the hypothesize-test as well-known approaches to evaluating a proposed argue [31] should be conducted. In this regard, method engineer should select a number of software development organizations to carry out the test so that they have categorized into two groups. One group constructs a project-specific service-oriented software development methodology via proposed process patterns while the other group does not use our process patterns during methodology construction and consequently does not use process patterns at their organizations. The non-parametric tests statistics, for instance Wilcoxon signed-rank test [32] helps to measure and compare the quality of software developed by the two groups. If the difference in quality of software development is significant, applicability and suitability of our proposed patterns are attested. Having said that, performing such a test is very expensive, time-consuming, and out of scope of our current report in this research

V. CONCLUSION AND FUTURE WORK

We proposed a set of comprehensive process patterns called SOPP that need to be engineered to form a situational service-oriented methodology. The proposed patterns were extracted from ten well-known service-oriented methodologies. SOPP presents service-oriented development process knowledge in a hierarchically structured and well defined way so that they can be used as reusable method chunks in SME approaches. To be more practical, a software development organization can use SOPP to tailor a development process according to the characteristics of a project. In this regards, we evaluated applicability of SOPP through a real-world case study.

While the aim of the research has been to present a high-level and abstract view of service-oriented development hence, we have focuses on WHAT should be done for service-oriented development in general instead of HOW this should be done. We are fully aware that the process patterns presented herein, focused on phases and stages patterns. At the present work, we are preparing detail definition of the task process patterns in order to publish them as method plug-in for EPFC environment. Additionally, the proposed patterns are enriched by introducing service-oriented *process anti-patterns* in which using a perfectly good patterns generates decidedly negative consequences in wrong contexts. We are planning to further improve SOPP using the experiences of its applications in real industry projects.

TABLE II. Utilized techniques for task patterns

| Task Patterns | Utilized Techniques | Requirements | Type |
|---|---|---|---|
| Elicit Requirements | Interviewing, questionnaire and checklist, brainstorming, sketching, textual analysis and storyboarding | #R9 | D |
| Priority of Requirements | MoSCoW Rules [29] | #R9 | D |
| Analyze Feasibility | EGT | #R9 | D |
| Validate Requirements | Evolutionary Prototyping [29[ | #R9 | D |
| Specify Requirements | User story [29] | #R9 | D |
| Specify SLA | Fill a template in which level of quality of services such as availability, security and performance are formally defined | #R1, #R6 | O |
| Evaluate Readability for Migration to SOA | As prescribed in [8] | #R4, #R5 | M |
| Decompose Environment | EGT | | D |
| Identify Policies and Rules | EGT | | M |
| Estimate for Budget and Resource | EGT | | D |
| Define Scope of Project | EGT | | D |
| Define Project Plan | EGT | | D |
| Plan Transition | As prescribed in [5] | #R4, #R5 | O |
| Assess Risk | By continuous review of project plan and user's feedback | | D |
| Model As-Is Business Process | EGT | #R2, #R3 | M |
| Decompose Business Process | EGT | #R2, #R3 | M |
| Identify Process Owners | EGT | #R2, #R3 | M |
| Identify Business Rules | EGT | #R2, #R3 | M |
| Identify Process's Quality Attributes | EGT | #R2, #R3 | M |
| Improve Business Process (To-Be) | EGT | #R2, #R3 | M |
| Define Services | Combination of top-down, bottom-up and goal service modeling techniques as prescribed in [7,10] | | D |
| Specify Services Interface | modeling techniques as prescribed by Erl [10] | | D |
| Analyze Services Collaboration | As prescribed in [7] | | D |
| Design Database | Design entity relation diagram (ERD) | | D |
| Classify Services | As prescribed in [10] | | D |
| Evaluate Services | As prescribed in [7] | | D |
| Develop Architecture | Instantiation of stack-based architecture as proposed in [10] | #R8 | M |
| Analyze SOA Technical Feasibility | As prescribed in [10] | #R8 | M |
| Design Technology Infrastructure | EGT | #R8 | M |
| Address Service Quality Concerns | Security Patterns, Architectural Patterns | #R6, #R8 | M |
| Evaluate Alternative Architecture | EGT such as ATAM and CBAM techniques | #R6, #R8 | M |
| Choose Implementation Alternatives | EGT | | D |
| Implement Wrappers | As prescribed in [7,10] | #R4 | M |
| Implement Services | OOA/D techniques such as grammatical parsing CRC card modeling and so on | #R2 | M |
| Select Technologies and Tools | Existing techniques | | D |
| Discover Services | As prescribed in [10] | #R1 | M |
| Certificate Services | As prescribed in [8] | #R1, #R6 | M |
| Enable Transition Plan | As prescribed in [8] | #R4, #R5 | M |
| Develop Plan for Test | EGT | | D |
| Generate Test Cases | EGT | | D |
| Run Test Cases | EGT | | D |
| Fix Bugs | EGT | | D |
| Run Walkthrough | Code Refactoring [29] | | D |
| Test Service Interfaces | EGT | | D |
| Publish Services | As prescribed in [8,12] | | D |
| Test Orchestrations / Choreographies Of services | EGT | | D |
| Add to Services Repository | As prescribed in [8,12] | | D |
| Prepare Infrastructure | EGT | | D |
| Monitor SLA | As prescribed in [7,8,12] | #R1, #R6 | M |
| Fix Defects | EGT | | D |

M= Mandatory, originated directly from requirements  D = Derived indirectly from initial requirements, or inevitable in any situation
N = Not required for this situation  EGT = Exiting general techniques